\documentclass[12pt]{article}
\pdfoutput=1
\usepackage{jhep-mod}
\usepackage{bm}
\usepackage{amssymb}
\usepackage{pifont}
\usepackage{slashed} 
\usepackage{slantsc} 
\usepackage[applemac]{inputenc}
\title{Bounding the greybody factors for scalar excitations of the Kerr--Newman spacetime}
\author[a]{Petarpa Boonserm,}
\affiliation[a]{Department of Mathematics and Computer Science, Chulalongkorn University, \\
Bangkok 10330, Thailand}
\emailAdd{petarpa.boonserm@gmail.com}
\author[b]{Tritos Ngampitipan,}
\affiliation[b]{Department of Physics, Chulalongkorn University, Bangkok 10330, Thailand}
\emailAdd{tritos.ngampitipan@gmail.com}
\author[c]{{\sf and} Matt Visser\,}
\affiliation[c]{School of Mathematics, Statistics, and Operations Research, \\
Victoria University of Wellington, PO Box 600, Wellington 6140, New Zealand}
\emailAdd{matt.visser@msor.vuw.ac.nz}

\abstract{
Finding exact solutions for black-hole greybody factors is generically impractical; typically one resorts either to making semi-analytic or numerical estimates, or alternatively to deriving rigorous analytic bounds. Indeed, rigorous bounds have already been established for the greybody factors of Schwarzschild and Riessner--Nordstr\"om black holes, and more generally for those of arbitrary static spherically symmetric asymptotically flat black holes.  Adding rotation to the problem greatly increases the level of difficulty, both for purely technical reasons (the Kerr or Kerr--Newman black holes are generally much more difficult to work with than the Schwarzschild or Reissner--Nordstr\"om black holes), but also at a  \emph{conceptual} level (due to the generic presence of super-radiant modes).  In the current article we analyze bounds on the greybody factors for scalar excitations of the Kerr--Newman geometry in some detail, first for zero-angular-momentum modes, then for the non-super-radiant modes, and finally for the super-radiant modes.

\bigskip
\noindent
3 January 2014; \LaTeX-ed \today
}
\begin{document}
\maketitle
\def\R{{\mathbb{R}}}
\def\d{{\mathrm{d}}}
\def\sech{{\mathrm{sech}}}
\def\O{{\mathcal{O}}}
\def\Dirac{{\slashed D }}
\providecommand{\ceil}[1]{\left \lceil #1 \right \rceil }
\section{Introduction}
\label{S:Introduction}

The spacetime geometry of a black hole, in the region that interpolates between the horizon and spatial infinity, (the domain of outer communication), generically acts as a potential barrier that partially reflects both ingoing and outgoing excitations. (See for instance~\cite{Page:1976a, Page:1976b, Bekenstein:1977, Escobedo:2008}.) In the case of outgoing excitations (Hawking quanta) the resulting transmission probabilities are called ``greybody factors". Calculation of these greybody factors, when practical, is based on analyzing the excitations in terms of a Regge--Wheeler equation, (or closely related variant thereof), which in the non-super-radiant case reduces the problem to a one-dimensional barrier-penetration problem. 

Even then, finding exact solutions is mostly impractical, and one typically resorts either to making semi-analytic or numerical estimates, or to deriving rigorous analytic bounds. Indeed, rigorous bounds have already been established for the greybody factors of the Schwarzschild~\cite{Schwarzschild} and Riessner--Nordstr\"om~\cite{Tritos:2012, Tritos:2013} black holes, and more generally for arbitrary static spherically symmetric asymptotically flat black holes~\cite{Static-Spherically-Symmetric}. Some preliminary work on the Kerr--Newman spacetime is presented in reference~\cite{Tritos:Singapore}. Some of the new issues raised in dealing with rotating black holes are purely technical --- the specific form of the metric is much more complicated. But there are new conceptual issues to deal with as well --- the presence of super-radiant modes now adding extra  conceptual overhead. 

The technique we are using to derive rigorous bounds on the greybody factors is a technique of general applicability to bounding transmission probabilities for one-dimensional barrier penetration problems.  First developed in reference~\cite{bounds}, this quite general technique has subsequently been extended in several different ways~\cite{miller-good, bogoliubov, analytic, shabat-zakharov}, before then being specifically applied to the analysis of black-hole greybody factors in references~\cite{Schwarzschild, Tritos:2012, Tritos:2013, Static-Spherically-Symmetric, Tritos:Singapore}.  In the current article we shall analyze bounds on the greybody factors for scalar excitations of the Kerr--Newman geometry in some detail, first for the zero-angular-momentum $m=0$ mode, secondly for generic  non-super-radiant modes,  and finally for the super-radiant modes. 

\section{Regge--Wheeler equation}\label{S:RW}

The Regge--Wheeler equation for the Kerr--Newman geometry is considerably more complicated than that for non-rotating spacetimes. Particularly useful recent references are~\cite{Kokkotas:2010, Konoplya:2011, Konoplya:2013}, though a wealth of other relevant material is also available~\cite{Kokkotas:1999, Ferrari:2003, Samuelsson:2006, Ferrari:2011}. 
Begin by writing the Kerr--Newman geometry in the form~\cite{Newman:1965a, Newman:1965b}
\begin{equation}
\d s^2 = -{\Delta\over\Sigma}\left(\d t - a\sin^2\theta\;\d\phi\right)^2 
+{\sin^2\theta\over\Sigma}\left[a\,\d t - (r^2+a^2) \;\d \phi\right]^2
+ {\Sigma\over\Delta}\; \d r^2 + \Sigma\; \d \theta^2,
\end{equation}
where
\begin{equation}
\Delta = r^2 -2Mr +a^2+Q^2 = (r-r_+)(r-r_-);  \qquad \Sigma = r^2 + a^2\sin^2\theta.
\end{equation}
Here $M$ is the mass of the black hole, $J = Ma$ is its angular momentum, and $Q$ is its charge. The quantities $r_\pm$ denote the locations of the inner and outer horizons. 
Setting $Q\to0$ gives the Kerr spacetime~\cite{Kerr:1963, Kerr:book, Kerr:survey}.
Now consider a massless electrically neutral minimally coupled scalar field.  (Adding mass and electric charge to the scalar field is not intrinsically difficult~\cite{Kokkotas:2010}, but is somewhat tedious, so we shall not do so for now.) 

\subsection{Spheroidal harmonics}\label{SS:Spheroidal}

It is a standard result, see Carter~\cite{Carter:1968}, that one can then use separation of variables to consider field modes of the form
\begin{equation}
\Psi(r,\theta,\phi,t) =  {R_{\ell m}(r) \, S_{\ell m}(\theta) \, \exp(-i\omega t  +i m \phi)\over\sqrt{r^2+a^2}}.
\end{equation}
It is now a standard but quite tedious computation to verify that the ``spheroidal harmonics'' $S_{\ell m}(\theta) \, e^{im\phi}$ generalize the usual ``spherical harmonics'' $Y_{\ell m}(\theta,\phi)$, and satisfy the differential equation:
\begin{equation}
\left\{ {1\over\sin\theta} {\d\over\d\theta}\left[\sin\theta {\d\over\d\theta} \right] 
- a^2\omega^2\sin^2\theta - {m^2\over\sin^2\theta} 
+2ma\omega +\lambda_{\ell m}(a\omega)
\right\} S_{\ell m}(\theta) = 0. 
\end{equation}
(See for instance~\cite{Sasaki:2003} pp 26--27.)
Note this differential equation is independent of $M$ and $Q$, though it does indirectly depend on the angular momentum via the dimensionless combination $a\omega = (J/M)\omega$. 
Here the separation constant $\lambda_{\ell m}(a\omega)$ generalizes the usual quantity $\ell(\ell+1)$ occurring for spherical harmonics, and in fact in the slow-rotation limit we have
\begin{equation}
\lambda_{\ell m}(a\omega) = \ell(\ell+1) - 2m\,a\omega  +\{H_{\ell+1, m} - H_{\ell m}\} \; (a\omega)^2 +\O[(a\omega)^3], 
\end{equation}
with
\begin{equation}
H_{\ell m} = {2\ell(\ell^2-m^2)\over 4\ell^2-1}. 
\end{equation}
Some useful background references are~\cite{Press:1973, Fackerell:1977, Shibata:1995,  Tagoshi:1996}. 
Note that since the differential operator is negative definite we automatically have the constraint that $\lambda_{\ell m}(a\omega)+2ma\omega\geq 0$. (To establish this, simply multiply the differential equation by $\sin^2\theta \, S_{\ell m}(\theta)$ and integrate by parts.)
In fact, re-writing the differential equation as
\begin{equation}
\left\{ {1\over\sin\theta} {\d\over\d\theta}\left[\sin\theta {\d\over\d\theta} \right] 
- \left( a\omega\sin\theta - {m\over\sin\theta}\right)^2 +\lambda_{\ell m}(a\omega)
\right\} S_{\ell m}(\theta) = 0, 
\end{equation}
we can also see that $\lambda_{\ell m}(a\omega)\geq 0$, an observation that will prove to be useful in the calculation below.
Furthermore, the differential equation for the $S_{\ell m}(\theta)$ can be \emph{explicitly solved} in terms of the confluent Heun functions. 
Unfortunately, this observation is less useful than one might hope, simply because despite valiant efforts not enough is yet known about the mathematical properties of Heun functions~\cite{Heun:1, Heun:2, Heun:3, Heun:4}. 

\subsection{Effective potential}\label{SS:effective}

With these preliminaries out of the way, it is now straightforward to write down the Regge--Wheeler equation for the radial modes~\cite{Kokkotas:2010}
\begin{equation}
\left\{ {\d^2\over\d r_*^2} - U_{\ell m}(r) \right\} R_{\ell m}(r) = 0.
\end{equation}
Here we use the ``tortoise coordinate'' defined by
\begin{equation}
\d r_* = {r^2+a^2\over\Delta} \; \d r=  {r^2+a^2\over(r-r_+)(r-r_-)} \; \d r.
\end{equation}
Explicitly
\begin{equation}
r_* = r + {a^2+r_+^2\over r_+-r_-} \; \ln(r-r_+) - {a^2+r_-^2\over r_+-r_-} \; \ln(r-r_-).
\end{equation}
Thus $r_*$ runs from $+\infty$ at spatial infinity to $-\infty$ at the outer horizon, located  at $r= r_+$. This region, the ``domain of outer communication'', is the only part of the spacetime geometry relevant for current purposes. 
The ``effective potential'' $U_{\ell m}(r) $ is:
\begin{equation}
U_{\ell m}(r) = {\Delta\over(r^2+a^2)^2} \left( \lambda_{\ell m}(a\omega) + {(r\Delta)'\over  r^2+a^2} -  {3 r^2 \Delta\over (r^2+a^2)^2} \right)   - \left(\omega - {ma\over r^2+a^2}\right)^2.
\end{equation}
For calculational purpose it is now useful to define quantities
\begin{equation}
\varpi = {a\over a^2+r^2}, \qquad\hbox{and more specifically,} \qquad \Omega_+= {a\over a^2+r_+^2}. 
\end{equation}
Here $\varpi(r)$ is (perhaps somewhat vaguely) related to frame dragging, while $\Omega_+$ is the angular velocity of the event horizon. 
We can now write
\begin{equation}
U_{\ell m}(r) = V_{\ell m}(r) - \left(\omega - {m\varpi}\right)^2,
\end{equation}
with
\begin{equation}
V_{\ell m}(r) =  {\Delta\over(r^2+a^2)^2} \left\{ \lambda_{\ell m}(a\omega) + W_{MQJ}(r)\right\}.
\end{equation}
Here we have separated out the quantity 
\begin{equation}
W_{MQJ}(r)= {(r\Delta)'\over  r^2+a^2} -  {3 r^2 \Delta\over (r^2+a^2)^2}, 
\end{equation}
which depends only on the spacetime geometry, not on the multipole $(\ell m)$ under consideration.
This definition of $V_{\ell m}(r)$ is now as close as possible to our earlier usage in references~\cite{Schwarzschild, Tritos:2012, Tritos:2013, Static-Spherically-Symmetric}, and to the general (non-relativistic quantum mechanical) analyses of references~\cite{bounds, miller-good, bogoliubov, analytic, shabat-zakharov}. 

\subsection{Positivity properties}\label{SS:positivity}

We have already seen that the separation constant $\lambda_{\ell m}(a\omega)$ is positive. More subtly the quantity $W_{MQJ}(r)$ is also positive. (This result depends implicitly on the Einstein equations and the resulting special properties of the Kerr--Newman spacetime.) 

To check the positivity of $W_{MQJ}(r)$, we write 
\begin{equation}
\Delta = (r-r_+)(r-r_-); \qquad r_+ + r_- = 2M; \qquad r_+ r_- = a^2 + Q^2.
\end{equation}
In particular note that
\begin{equation}
0 \leq {a^2\over r_+} \leq r_- \leq r_+,
\qquad
\hbox{and}
\qquad
0 \leq {Q^2\over r_+} \leq r_- \leq r_+.
\end{equation}
Furthermore
\begin{equation}
a \leq M;\qquad |Q| \leq M.
\end{equation}
Now consider
\begin{eqnarray}
(r \Delta)' &=& [r(r-r_+)(r-r_-)]' \nonumber\\
&=& (r-r_+)(r-r_-) +r(r-r_+)+r(r-r_-) \nonumber\\
&=& 3 r^2 - 2 r(r_++r_-)+r_+r_-.
\end{eqnarray}
Then 
\begin{eqnarray}
W_{MQJ}(r) &\propto&  (r \Delta)'  (r^2+a^2)-3 r^2 \Delta \nonumber\\
&=& (3 r^2 - 2 r(r_++r_-)+r_+r_-)(r^2+a^2)-3r^2(r-r_+)(r-r_-) \nonumber\\
&=& [0] r^4 + [-2(r_++r_-)+3(r_++r_-) ]r^3 + [3a^2+r_+r_--3r_+r_- ]r^2 \nonumber\\
&& \qquad + [-2a^2(r_++r_-)]r +[a^2 r_+r_-] r^0 \nonumber\\
&=& (r_++r_-)r^3 + [3a^2-2r_+r_- ]r^2 -2a^2(r_++r_-)r +a^2 r_+r_-\nonumber\\
&=& r^2(r r_+ + r r_- - 2r_+ r_-) + a^2r(2r-r_+-r_-) + a^2\Delta \nonumber\\
&\geq& 0. 
\end{eqnarray}
Here in the penultimate line all three terms are manifestly positive outside the outer horizon (for $r\geq r_+$). 

Furthermore $\lim_{r\to\infty} W_{MQJ} =0$ and $W_{MQJ}(r_+) = r_+(r_+-r_-)/(r_+^2+a^2)$. Thence we see that $V_{\ell m}\to0$ both at the outer horizon $r_+$ and at spatial infinity.

\subsection{Super-radiance}\label{SS:super-radiance}

It is the trailing term in the effective potential, the $\left(\omega - m\varpi\right)^2$ term, that is responsible for the qualitatively new phenomenon of super-radiance, which never occurs in ordinary non-relativistic quantum mechanics. The reason for this is that the Schr\"odinger equation is first-order in time derivatives, so the effective potential for Schr\"odinger-like barrier-penetration problems is generically of the form
\begin{equation}
U(r)   = V(r)  - \omega.
\end{equation}
In contrast, for problems based on the Klein--Gordon equation (second-order in time derivatives) the qualitative structure of the effective potential is
\begin{equation}
U(r) = V(r) - (\omega - m\varpi)^2.
\end{equation}
We shall soon see that it is when the quantity $\omega-m\varpi$ changes sign that the possibility of super-radiance arises. (See for instance the general discussion by Richartz \emph{et al}~\cite{Richartz:2009}.)   In the current set-up super-radiance is related to the rotation of the black hole, but if the scalar field additionally carries electric charge there is another contribution to $\varpi$ coming from the electrostatic potential, and so a separate route to super-radiance~\cite{Kokkotas:2010, Richartz:2009}. 

While the Dirac equation, being first-order in both space and time, might seem to side-step this phenomenon, it is a standard result that iterating the Dirac differential operator twice produces a Klein--Gordon-like differential equation. In terms of the Dirac matrices we have:
\begin{equation}
\Dirac^2 = 2 (\nabla-iqA)^2 + q F_{ab} \; [\gamma^a, \gamma^b].
\end{equation}
So, once one factors out the spinorial components, and concentrates attention on the second-order differential equation for the amplitude of the Dirac field, even the Klein paradox for charged relativistic fermions can be put into this framework. It is the trailing $(\omega - m\varpi)^2$ term, and more specifically the change in sign of $\omega - m\varpi$, that is the harbinger of super-radiance. Indeed, assuming $\varpi$ is monotonic (which it certainly is in the situations we shall be interested in) let us define the quantity $m_*=\omega/\Omega_+$. Then:
\begin{itemize}
\item the modes $m< {m_*}$ are not super-radiant;
\item the modes $m \geq {m_*}$ are super-radiant.
\end{itemize}
We shall soon see much more detail regarding the super-radiance phenomenon in the subsequent discussion.

\section{Non-super-radiant modes ($m < {m_*}$) }\label{S:nonsuper}

It is convenient to split the discussion of the non-super-radiant modes into three sub-cases:
\begin{itemize}
\item $m=0$, zero-angular-momentum modes;
\item $m<0$, negative-angular-momentum modes;
\item $m\in(0,{m_*})$, low-lying positive-angular-momentum modes. 
\end{itemize}

\subsection{Zero-angular-momentum modes ($m=0$)}\label{SS:Zero}
This sub-case is both particularly simple, and is in many ways a guiding template for all the other cases. 
Some preliminary work on these zero-angular-momentum modes in the  Kerr--Newman spacetime is presented in reference~\cite{Tritos:Singapore}. 
We note that from reference~\cite{bounds} pp. 427--428 we have the very generic bound:
\begin{equation}
T_{\ell m} \geq \sech^2\left\{  \int_{-\infty}^{+\infty} {\sqrt{ [h'(r)]^2 +[U_{\ell m}(r)+h(r)^2]^2}\over 2 h(r)} \; \d r_* \right\}; \qquad \forall h(r) > 0.
\end{equation}
Note that we need $h(r)>0$ everywhere in order for this bound to hold.
Suppose we set  $m=0$, then
 \begin{equation}
U_{\ell,m=0}(r) =  - \omega^2 + 
{\Delta\over(r^2+a^2)^2} \left[ \lambda_{\ell,m=0} + W_{MQJ}(r) \right].
\end{equation}
Now choose $h(r) =\omega > 0$, and change the integration variable from $\d r_*$ to $\d r$, so that
\begin{equation}
T_{\ell,m=0} \geq \sech^2\left\{ {1\over2\omega}\int_{r_+}^{+\infty} \left|  
{1\over(r^2+a^2)} \left[ \lambda_{\ell,m=0} +W_{MQJ}(r) \right]  
 \right| \; \d r   \right\}.
\end{equation}
(This corresponds to the Case I bound of reference~\cite{bounds}.)
As long as $\lambda_{\ell m}$ and $W_{MQJ}(r)$ are always positive (and we have already checked that above) we can dispense with the absolute value symbols and write
\begin{equation}
T_{\ell,m=0} \geq \sech^2\left\{ {1\over2\omega}\int_{r_+}^{+\infty} 
{1\over(r^2+a^2)} \left[ \lambda_{\ell,m=0} +W_{MQJ}(r) \right]   \; \d r   \right\}.
\end{equation}
This now decouples the problem to considering two integrals, each of which can be explicitly evaluated in closed form.

\paragraph{First integral:} We note that
\begin{equation}
\int_{r_+}^{+\infty}  { \lambda_{\ell,m=0}\over(r^2+a^2)} \d r = \lambda_{\ell,m=0}(a\omega) \;\; {\arctan(a/r_+)\over a}.
\end{equation}
This quantity is independent of $M$ and $Q$. 

\paragraph{Second integral:}
When it comes to evaluating the integral involving $W_{MQJ}$ it is best to define the dimensionless quantity
\begin{equation}
K_{MQJ} = r_+ \int_{r_+}^{+\infty} 
{W_{MQJ}\over(r^2+a^2)}
\d r
=
r_+ \int_{r_+}^{+\infty} 
{1\over(r^2+a^2)} \left( {(r\Delta)'\over  r^2+a^2} -  {3 r^2 \Delta\over (r^2+a^2)^2} \right)  
\d r.
\end{equation}
To evaluate this the best trick is to integrate by parts:
\begin{equation}
K_{MQJ} = r_+ 
\int_{r_+}^{+\infty} 
\left( -(r\Delta) [(r^2+a^2)^{-2}]' -  {3 r^2 \Delta\over (r^2+a^2)^3} \right) 
\d r.
\end{equation}
(Note that the boundary terms vanish). This then equals:
\begin{equation}
K_{MQJ} = r_+ \int_{r_+}^{+\infty} 
\left(   {(4-3) r^2 \Delta\over (r^2+a^2)^3} \right) 
\d r 
=
r_+  \int_{r_+}^{+\infty} 
\left(   { r^2 \Delta\over (r^2+a^2)^3} \right) 
\d r.
\end{equation}
So finally
\begin{equation}
K_{MQJ} = {r_+\over 8}  { (r_+^2+a^2)(3a^2+r_+r_- )\arctan(a/r_+)    + a(a^2[r_+-2r_-] -r_+^2 r_-)   \over a^3(r_+^2+a^2)}.
\end{equation}
This dimensionless quantity is independent of the parameters characterizing the scalar mode $(\ell, m, \omega)$, and depends only on the parameters characterizing the spacetime geometry $(a,r_+,r_-)$, which in turn implicitly depend only on $(M,Q,J)$.

\paragraph{Consistency check:}
If you look carefully this quantity $K_{MQJ}$ does have a finite limit as $a\to 0$, as it should do to be consistent with the physics of the Reissner--Nordstr\"om spacetime. (The limit is a little tricky.) We can recast $K_{MQJ}$ as
\begin{equation}
K_{MQJ} = {3\over8} {\arctan(a/r_+)    \over a/r_+} + {r_+^2r_-\over 8} {([r_+^2+a^2]\arctan(a/r_+)-  a r_+  )\over  a^3(r_+^2+a^2)}
+{1\over 8}{r_+(3 a+r_+-2r_-)\over r_+^2+a^2} ,
\end{equation}
with limit
\begin{equation}
\to {3\over 8}  + {1\over12} {r_-\over r_+}  + {1\over8}{r_+-2r_-\over r_+} 
= {1\over24}{ 9r_+ + 2r_- +3 r_+ - 6 r_-\over r_+} 
={3r_+-r_-\over 6r_+}.
\end{equation}

\paragraph{Final result:}
Collecting terms, we can write the bound on the transmission probability as
\begin{equation}
T_{\ell,m=0} \geq \sech^2 \left[  {I_{\ell,m=0}\over2 r_+ \omega} \right],
\end{equation}
with
\begin{equation}
I_{\ell,m=0} =  \lambda_{\ell,m=0}(a\omega)  \; {\arctan(a/r_+)\over a/r_+}    +  K_{MQJ}.
\end{equation}
This cleanly separates out the mode dependence $(\ell m)$ from the purely geometrical piece $K_{MQJ}$. 
Note $I_{\ell,m=0}$ is now a dimensionless number that depends only dimensionless ratios such as $a/r_+$ and $r_-/r_+$, and implicitly (via $\lambda_{\ell,m=0}$) on $\ell$ and $a\omega$. 
In view of the known slow rotation expansion for $\lambda_{\ell,m=0}(a\omega) $ we know that
\begin{equation}
I_{\ell,m=0}(\omega\to0) =  \ell(\ell+1)  \; {\arctan(a/r_+)\over a/r_+}    +  K_{MQJ}.
\end{equation}
So at low frequencies the transmission bound is dominated by the $1/\omega$ pole in the argument of the hyperbolic secant function. 
If we wish to be very explicit we can write
\begin{eqnarray}
I_{\ell,m=0} &=& \left(\lambda_{\ell,m=0}(a\omega)+ {3\over 8}\right) \; {\arctan(a/r_+)\over a/r_+}  
\\
&&
+  {r_+r_-\over 8} \; {r_+([r_+^2+a^2]\arctan(a/r_+)-  a r_+  )\over  a^3(r_+^2+a^2)}
+{1\over 8}{r_+(3 a+r_+-2r_-)\over r_+^2+a^2}.
\nonumber
\end{eqnarray}
There are certainly other ways of re-writing this quantity, but this version is sufficient for exhibiting key aspects of the physics. 

\subsection{Non-zero-angular-momentum modes ($m\neq0$)}\label{S:Non-zero}

What if anything can we do once $m\neq 0$? Recall the basic result
\begin{equation}
T_{\ell m} \geq \sech^2\left\{  \int_{-\infty}^{+\infty} {\sqrt{ [h'(r)]^2 +[U_{\ell m}(r)+h(r)^2]^2}\over 2 h(r)} \; \d r_* \right\}; \qquad \forall h(r) > 0.
\end{equation}
Now by the triangle inequality we certainly have
\begin{equation}
T_{\ell m} \geq \sech^2\left\{  {1\over2} \int_{-\infty}^{+\infty} \left| h'\over h\right| \d r_* 
+  {1\over2} \int_{-\infty}^{+\infty}{|U_{\ell m}(r)+h(r)^2|\over 2 h(r)} \; \d r_* \right\}; \qquad \forall h(r) > 0.
\end{equation}
We are now free to pick $h(r)$ so that it is monotone, $h'(r) > 0$ or $h'(r)<0$. Then subject to this condition
\begin{equation}
T_{\ell m} \geq \sech^2\left\{  {1\over2}\left| \ln\left[ h(\infty)\over h(-\infty)\right] \right|  
+  {1\over2} \int_{-\infty}^{+\infty}{|U_{\ell m}(r)+h(r)^2|\over 2 h(r)} \; \d r_* \right\}; \qquad \forall h(r) > 0.
\end{equation}
Apply this general result to our specific situation
\begin{equation}
U_{\ell m}(r) = V_{\ell m}   - \left(\omega - {m\varpi}\right)^2,
\end{equation}
by choosing
\begin{equation}
h(r) = \omega - {m\varpi}.
\end{equation}
(This construction is now as close as one can get to the Case I bound of reference~\cite{bounds}.)
Note this choice for $h(r)$ is, since $\varpi = a/(a^2+r^2)$, always monotonic as a function of $r$. In contrast, (remember that $\omega>0$ and $a>0$), we see that this $h(r)$ is positive throughout the domain of outer communication \emph{if and only if} $\omega > m \Omega_+$, which is completely equivalent to $m< \omega/\Omega_+$, or $m<m_*$.  
This is easily recognized as the quite standard condition that 
the mode does \emph{not} suffer from super-radiant instability.   
Let us now see where we can go with this.

\subsubsection{Negative-angular-momentum modes ($m<0$)}
First note that in this situation, for the specific function $h(r)$ chosen above, we have
\begin{equation}
 {h(\infty)\over h(-\infty)} = {\omega\over \omega-m\Omega_+} = {1\over 1 - m\Omega_+/\omega} < 1.
 \end{equation}  
Then
\begin{equation}
 {1\over2}\left| \ln\left[ h(\infty)\over h(-\infty)\right] \right|   = {1\over2} \ln(1 - m\Omega_+/\omega).
\end{equation}
Also in this case we have $  \omega-m\Omega_+ >  h(r) > \omega$, so
\begin{equation}
 \int_{-\infty}^{+\infty}{|U_{\ell m}(r)+h(r)^2|\over 2 h(r)} \; \d r_* 
 =  
   \int_{-\infty}^{+\infty}{|V_{\ell m}|\over 2 h(r)} \; \d r_* 
  <  
  \int_{-\infty}^{+\infty}{V_{\ell m}\over 2 \omega} \; \d r_*.
 \end{equation}
 Then 
 \begin{equation}
 T_{\ell,m<0} \geq \sech^2\left\{  {1\over2} \ln(1 - m\Omega_+/\omega) 
 +  \int_{-\infty}^{+\infty}{V_{\ell,m<0}\over 2 \omega} \; \d r_*.
\right\}
 \end{equation}
 But that last integral is almost identical to that we performed for $m=0$, the only change being the replacement $\lambda_{\ell,m=0} \to \lambda_{\ell, m<0}$. Therefore
 \begin{equation}
 T_{\ell,m<0} \geq \sech^2\left\{  {1\over2} \ln(1 - m\Omega_+/\omega) +  {I_{\ell, m<0}\over2 r_+ \; \omega} \right\},
\end{equation}
where in comparison we previously had
 \begin{equation}
T_{\ell,m=0} \geq \sech^2 \left\{  { I_{\ell,m=0}  \over2 r_+ \omega}\right\}.
\end{equation}
Explicitly
\begin{equation}
I_{\ell m} =  \lambda_{\ell m}(a\omega)  \; {\arctan(a/r_+)\over a/r_+}    +  K_{MQJ},
\end{equation}
and
 \begin{equation}
 T_{\ell,m<0} \geq \sech^2\left\{  {1\over2} \ln(1 - m\Omega_+/\omega) 
 +  {  \lambda_{\ell m}(a\omega)  \;\; {\textstyle\arctan(a/r_+)\over\textstyle a/r_+}    +  K_{MQJ}\over2 r_+ \; \omega} \right\}.
\end{equation}
Note that for $m<0$ we have $-m \leq \ell$, so we could also write the weaker (but perhaps slightly simpler) bound 
  \begin{equation}
 T_{\ell,m<0} \geq \sech^2\left\{  {1\over2} \ln(1 +\ell\Omega_+/\omega) +  {I_{\ell, m<0}\over2 r_+ \;\omega} \right\}.
\end{equation}
 
\subsubsection{Low-lying positive-angular-momentum modes ($m\in(0,{m_*})\,$)}
For this situation we first note that
\begin{equation}
 {h(\infty)\over h(-\infty)} = {\omega\over \omega-m\Omega_+} = {1\over 1 - m\Omega_+/\omega} > 1.
 \end{equation}  
Then we see
\begin{equation}
 {1\over2}\left| \ln\left[ h(\infty)\over h(-\infty)\right] \right|   = -{1\over2} \ln(1 - m\Omega_+/\omega).
\end{equation}
Also, in this case $  \omega-m\Omega_+ <  h(r) < \omega$, so
\begin{equation}
 \int_{-\infty}^{+\infty}{|U_{\ell m}(r)+h(r)^2|\over 2 h(r)} \; \d r_* 
  =
  \int_{-\infty}^{+\infty}{|V_{\ell,m>0}|\over 2 h(r)} \; \d r_* 
  <  
  \int_{-\infty}^{+\infty}{V_{\ell,m>0}\over 2 (\omega-m\Omega_+)} \; \d r_*.
 \end{equation}
 Then 
 \begin{equation}
 T_{\ell,m>0} \geq \sech^2\left\{  -{1\over2} \ln(1 - m\Omega_+/\omega) 
 +  {1\over2} \int_{-\infty}^{+\infty}{|V_{\ell,m>0}|\over (\omega-\Omega_+)} \; \d r_*.
\right\}
 \end{equation}
 But that remaining integral is qualitatively the same as that which we performed for the $m=0$ and $m<0$ cases, therefore
 \begin{equation}
 T_{\ell,m>0} \geq \sech^2\left\{  -{1\over2} \ln(1 - m\Omega_+/\omega) +  {I_{\ell,m>0}\over2 r_+ (\omega-m\Omega_+)}
 \right\},
\end{equation}
 where in comparison
 \begin{equation}
T_{\ell,m=0} \geq \sech^2 \left\{ {I_{\ell,m=0}\over2 r_+ \;\omega} \right\}.
\end{equation}
Explicitly
\begin{equation}
I_{\ell m} =  \lambda_{\ell m}(a\omega)  \; {\arctan(a/r_+)\over a/r_+}    +  K_{MQJ},
\end{equation}
and
 \begin{equation}
 T_{\ell,m>0} \geq \sech^2\left\{  -{1\over2} \ln(1 - m\Omega_+/\omega) 
 +
 { \lambda_{\ell m}(a\omega)  \; {\textstyle\arctan(a/r_+)\over\textstyle a/r_+}    +  K_{MQJ}\over2 r_+ (\omega-m\Omega_+)}
 \right\}.
\end{equation}
Note that for $m>0$ we have $m \leq \ell$, so we could also write the weaker (but perhaps slightly simpler) bound 
  \begin{equation}
 T_{\ell,m>0} \geq \sech^2\left\{  -{1\over2} \ln(1 -\ell\Omega_+/\omega) +  {I_{\ell m} \over2 r_+ (\omega-\ell\Omega_+)} \right\}.
\end{equation}

\subsection{Summary (non-super-radiant modes)}\label{S:Summary1}

Define
\begin{equation}
I_{\ell m} =  \lambda_{\ell m}(a\omega)  \; {\arctan(a/r_+)\over a/r_+}    +  K_{MQJ},
\end{equation}
where
\begin{equation}
K_{MQJ} = {r_+\over 8}  { (r_+^2+a^2)(3a^2+r_+r_- )\arctan(a/r_+)    + a(a^2[r_+-2r_-] -r_+^2 r_-)   \over a^3(r_+^2+a^2)}.
\end{equation}
Then for the non-super-radiant modes
\begin{equation}
 T_{\ell,m\leq0} \geq \sech^2\left\{  {1\over2} \ln(1 - m\Omega_+/\omega) +  {I_{\ell, m\leq0}\over2 r_+ \; \omega} \right\},
\end{equation}
and
 \begin{equation}
 T_{\ell,m\in(0,m_*)} \geq \sech^2\left\{  -{1\over2} \ln(1 - m\Omega_+/\omega) +  {I_{\ell, m>0}\over2 r_+ (\omega-m\Omega_+)} \right\}.
\end{equation}
These bounds can also be written as
\begin{equation}
 T_{\ell,m\leq0} \geq \sech^2\left\{  {1\over2} \ln(1 - m/m_*) +  {I_{\ell, m\leq0}\over2 r_+ \; \omega} \right\},
\end{equation}
and
 \begin{equation}
 T_{\ell,m\in(0,m_*)} \geq \sech^2\left\{  -{1\over2} \ln(1 - m/m_*) +  {I_{\ell, m>0}\over2 r_+ \; \omega\; (1-m/m_*)} \right\}.
\end{equation}
These are the best general bounds we have been able to establish for the non-super-radiant modes.

\section{Super-radiant modes ($m\geq {m_*}$)}\label{SS:super-radiant}

For the super-radiant modes we must be more careful. Inspection of the original derivation in reference~\cite{bounds} shows that fundamentally the analysis works by placing bounds on the Bogoliubov coefficients: 
\begin{equation}
|\alpha| \leq \cosh \oint \vartheta \;\d r; \qquad |\beta| \leq  \sinh \oint \vartheta \;\d r,
\end{equation}
where
\begin{equation}
\Theta= \oint \vartheta \;\d r =  \int_{-\infty}^{+\infty} {\sqrt{ [h'(r)]^2 +[U_{\ell m}(r)+h(r)^2]^2}\over 2 h(r)} \; \d r_*; \qquad \forall h(r) > 0.
\end{equation}
In the non-super-radiant case these constraints on the Bogoliubov coefficients quickly and directly  lead to a bound on the transmission coefficient $T = |\alpha|^{-2}$.  In counterpoint, in the super-radiant case the Bogoliubov coefficients also have an additional physical interpretation: The near-horizon quantum vacuum state now contains a nontrivial density of quanta when viewed from the region near spatial infinity~\cite{Richartz:2009}. The number of quanta per unit length in each mode is $n =  k\; |\beta|^2$, corresponding to an emission rate 
\begin{equation}
\Gamma = \omega \; |\beta|^2.
\end{equation}
Explicitly, the emission rate in each specific mode is bounded by
\begin{equation}
\Gamma_{\ell m}(\omega) \leq  \omega \sinh^2 \Theta,
\end{equation}
where
\begin{equation}
\Theta =  \int_{-\infty}^{+\infty} {\sqrt{ [h'(r)]^2 +[U_{\ell m}(r)+h(r)^2]^2}\over 2 h(r)} \; \d r_*; \qquad \forall h(r) > 0.
\end{equation}
The net result is that one is still interested in the same integral, but now under different conditions, and with an additional  physical interpretation. To be more explicit about this, note that
\begin{equation}
\Theta =  \int_{-\infty}^{+\infty} {\sqrt{ [h'(r)]^2 +[V_{\ell m}(r)-(\omega-m\varpi(r))^2+h(r)^2]^2}\over 2 h(r)} \; \d r_*; \quad \forall h(r) > 0.
\end{equation}
The art comes now in \emph{choosing} a specific $h(r)$ to in some sense optimize the bound, (either by making it a particularly tight bound, or by making it a particularly simple bound),  subject now to the condition that $\omega-m\varpi(r)$ is assumed to change sign at some finite value of $r$, and subject to the condition that one wants the integral to be finite, (implying in particular that the integrand should vanish both on the outer horizon and at spatial infinity). 

\noindent
Now the triangle inequality implies ($\forall h(r)>0$) that
\begin{equation}
\Theta \leq   \  {1\over2} \int_{-\infty}^{+\infty} {|h'(r)|\over h(r)} \; \d r_*
+ \int_{-\infty}^{+\infty} {|V_{\ell m}(r)-(\omega-m\varpi(r))^2+h(r)^2|\over 2 h(r)} \; \d r_*.
\end{equation}
Additionally we know that $V_{\ell m}\to 0$ at both the outer horizon and spatial infinity, so to keep the integral finite we need both $h(\infty)^2=\omega^2$ and $h(r_+)^2=(\omega-m\Omega_+)^2$.  
Based on this observation, it is now a good strategy to again use the triangle inequality to split the integral as follows 
\begin{equation}
\Theta \leq   \  {1\over2} \int_{-\infty}^{+\infty} {|h'(r)|\over h(r)} \; \d r_*
+ \int_{-\infty}^{+\infty} {V_{\ell m}(r)\over 2 h(r)} \; \d r_*
+ \int_{-\infty}^{+\infty} {|h(r)^2-(\omega-m\varpi(r))^2|\over 2 h(r)} \; \d r_*.
\end{equation}
Now split the 
super-radiant modes into two sub-cases
depending on the relative sizes of $\omega^2$ and $(\omega - m \Omega_+)^2$.
But note that in the super-radiant regime $\omega^2 = (\omega - m \Omega_+)^2$ when  $m = 2 \omega/\Omega_+ = 2m_*$. This suggests splitting the super-radiant regime into two distinct sub-cases:
\begin{itemize}
\item $ m \in  [m_*,    2 m_*)$.
\item $ m \in [2 m_*, \infty)$.
\end{itemize}

\subsection{Low-lying super-radiant modes ($m\in[m_*, 2 m_*)\,$)}\label{SS:low-super-radiant}
In this region  we have $\omega^2 > (\omega - m \Omega_+)^2$ and so we could take:
\begin{equation}
 h(r) =  \max\left\{ \omega- {m a\over(a^2+r^2)} ,  m \Omega_+ - \omega \right\}.
\end{equation}  
This quantity is positive and monotone decreasing as we move from spatial infinity to the horizon,
and becomes a flat horizontal line near the horizon. Note that by construction $h(r) \geq m \Omega_+ - \omega$ everywhere.
First, from the definition of $h(r)$, in this situation we have
\begin{equation}
\int_{-\infty}^{+\infty} {|h'(r)|\over h(r)} \; \d r_* = |\ln h(r)|^\infty_{r_+} = \ln\left({\omega\over m\Omega_+-\omega}\right) = - \ln(m/m_*-1).
\end{equation}
Second
\begin{equation}
 \int_{-\infty}^{+\infty} {V_{\ell m}(r)\over 2 h(r)} \; \d r_* 
 \leq  \int_{-\infty}^{+\infty} {V_{\ell m}(r)\over 2 (m\Omega_+-\omega)} = {I_{\ell m}\over 2 (m\Omega_+-\omega)} = {I_{\ell m}\over 2 \omega (m/m_*-1)} ,
\end{equation}
where the $I_{\ell m}$ integral is the same quantity we have considered several times before.
Finally, the remaining integral to be performed is 
\begin{equation}
J_{m}^\mathrm{low} = \int_{-\infty}^{+\infty} {h(r)^2-(\omega-m\varpi(r))^2\over 2 h(r)} \; \d r_*,
\end{equation}
with the integrand being both independent of $\ell$, and carefully chosen to be zero over much of the relevant range. Indeed, unwrapping all of the definitions,  we are interested in
\begin{equation}
J_{m}^\mathrm{low}  = \int_{r_+}^{r_0} {(\omega-m\Omega_+)^2-(\omega-m\varpi(r))^2\over 2 (m\Omega_+-\omega)} \;\; {r^2+a^2\over\Delta}\; \d r.
\end{equation}
The upper limit of integration $r_0$ is defined by 
\begin{equation}
m[\Omega_++a/(a^2+r_0^2)]=2\omega,
\end{equation}
that is, by 
\begin{equation}
r_0^2-r_+^2 =  {2(m-m_*)\over2m_*-m}\;    (r_+^2+a^2).
\end{equation}
Explicitly
\begin{equation}
r_0 =  \sqrt{ r_+^2 + {2(m-m_*)\over2m_*-m}\;    (r_+^2+a^2)}.
\end{equation}
Note $r_0>r_+$ for $m\in[m_*, 2 m_*)$. 
Then
\begin{equation}
J_{m}^\mathrm{low}  = {m \over 2 (\omega-\Omega_+)}  \int_{r_+}^{r_0} (\Omega_+ -\varpi) (2\omega-m\varpi(r)-m\Omega_+)\;\; {r^2+a^2\over\Delta}\; \d r.
\end{equation}
But over the relevant domain $0 \leq (2\omega-m\varpi(r)-m\Omega_+ \leq 2(\omega-m\Omega_+)$, therefore
\begin{equation}
J_{m}^\mathrm{low}  \leq m \int_{r_+}^{r_0} (\Omega_+ -\varpi) \;\; {r^2+a^2\over\Delta}\; \d r.
\end{equation}
The remaining integral is now simple and manifestly finite.
\begin{equation}
J_{m}^\mathrm{low}  \leq m \int_{r_+}^{r_0} (\Omega_+ -\varpi) \;\; {r^2+a^2\over\Delta}\; \d r 
= {ma\over r_+^2+a^2} \int_{r_+}^{r_0} {r-r_+\over r-r_-}\; \d r 
\end{equation}
(In fact we could have evaluated $J_m^\mathrm{low}$ exactly, but given the other approximations being made in deriving the bounds, there is no real point in doing so.) 
Assembling the pieces we have:
 \begin{equation}
 T_{\ell,m\in[m_*,2m_*)} \geq \sech^2\left\{  -{1\over2} \ln(m/m_*-1) +  {I_{\ell, m\in[m_*,2m_*)}\over2 r_+ \omega\,(m/m_*-1) } + J_m^\mathrm{low}\right\}.
\end{equation}
Furthermore:
 \begin{equation}
 \Gamma_{\ell,m\in[m_*,2m_*)} \leq \omega\; \sinh^2\left\{  -{1\over2} \ln(m/m_*-1) +  {I_{\ell, m\in[m_*,2m_*)}\over2 r_+ \omega\,(m/m_*-1) } + J_m^\mathrm{low}\right\}.
\end{equation}

\subsection{Highly super-radiant modes ($m\geq 2 m_*$)}\label{SS:high-super-radiant}
In this region we have $(\omega - m \Omega_+)^2 > \omega^2$ and so we could take:
\begin{equation}
  h(r) =  \max\left\{  {m a\over(a^2+r^2)} -\omega ,  \omega \right\}
\end{equation}
This is now both positive and monotone decreasing as we move from the horizon to spatial infinity,
and becomes a flat horizontal line near spatial infinity. Note $h(r) \geq \omega$ everywhere.
First, from the definition of $h(r)$, in this situation we have
\begin{equation}
\int_{-\infty}^{+\infty} {|h'(r)|\over h(r)} \; \d r_* = |\ln h(r)|^\infty_{r_+} = \ln\left({m\Omega_+-\omega\over\omega}\right) = \ln(m/m_*-1).
\end{equation}
Second
\begin{equation}
 \int_{-\infty}^{+\infty} {V_{\ell m}(r)\over 2 h(r)} \; \d r_* 
 \leq  \int_{-\infty}^{+\infty} {V_{\ell m}(r)\over 2 \omega} = {I_{\ell m}\over 2 \omega},
\end{equation}
where the $I_{\ell m}$ integral is the same quantity we have considered before.
Finally, the remaining integral is 
\begin{equation}
J_{m}^\mathrm{high} = \int_{-\infty}^{+\infty} {h(r)^2-(\omega-m\varpi(r))^2\over 2 h(r)} \; \d r_*,
\end{equation}
with the integrand being zero over much of the relevant range. Indeed we are now interested in
\begin{equation}
J_{m}^\mathrm{high} =  \int_{r_0}^{\infty} {\omega^2-(\omega-m\varpi(r))^2\over 2 \omega} \;\; {r^2+a^2\over\Delta}\; \d r,
\end{equation}
The lower limit of integration $r_0$ is now defined by $ma/(a^2+r_0^2)=2\omega$, that is, by 
\begin{equation}
r_0 = a \sqrt{{m\over2\omega a}-1}.
\end{equation}
Note that since $m \geq 2m_*$ we have
\begin{equation}
r_0 \geq a \sqrt{{m_*\over\omega a}-1}= a \sqrt{{a^2+r_+^2\over a^2}-1} = r_+,
\end{equation}
so we are safely outside (or possibly just on) the outer horizon. If $m>2m_*$ then $r_0>r_+$ and the integrand is manifestly finite over the entire range of interest, while falling of asymptotically as $ 1/r^2$, so the integral $J_{m}^\mathrm{high}$ is finite. If $m=2m_*$ so $r_0=r_+$, then both the numerator and denominator of the integrand to zero at the outer horizon, though the ratio is finite. So the integrand again remains finite over the entire range of interest, while falling of asymptotically as $ 1/r^2$, so the integral $J_{m}^\mathrm{high}$ is again finite. (In fact we can evaluate $J_m^\mathrm{low}$ exactly, but the result is algebraically messy, and given the other approximations being made in deriving the bounds, there is no real point in doing so.) 
Assembling the pieces we have:
 \begin{equation}
 T_{\ell,m\geq2m_*} \geq \sech^2\left\{  {1\over2} \ln(m/m_*-1) +  {I_{\ell, m\geq 2m_*}\over2 r_+ \omega } + J_m^\mathrm{high}\right\}.
\end{equation}
Furthermore:
 \begin{equation}
 \Gamma_{\ell,m\geq 2m_*} \leq \omega\; \sinh^2\left\{  {1\over2} \ln(m/m_*-1) +  {I_{\ell, m\geq2m_*}\over2 r_+ \omega } + J_m^\mathrm{high}\right\}.
\end{equation}

\bigskip
\subsection{Summary (super-radiant modes)}\label{S:Summary2}

Pulling the results for the low-lying and highly super-radiant modes together we see that for the transmission probabilities we have:
\begin{equation}
 T_{\ell,m\in[m_*,2m_*)} \geq \sech^2\left\{  -{1\over2} \ln(m/m_*-1) +  {I_{\ell, m\in[m_*,2m_*)}\over2 r_+ \omega\,(m/m_*-1) } + J_m^\mathrm{low}\right\}.
\end{equation}
 \begin{equation}
 T_{\ell,m\geq2m_*} \geq \sech^2\left\{  {1\over2} \ln(m/m_*-1) +  {I_{\ell, m\geq 2m_*}\over2 r_+ \omega } + J_m^\mathrm{high}\right\}.
\end{equation}
Furthermore for the super-radiant emission rates we have:
 \begin{equation}
 \Gamma_{\ell,m\in[m_*,2m_*)} \leq \omega\; \sinh^2\left\{  -{1\over2} \ln(m/m_*-1) +  {I_{\ell, m\in[m_*,2m_*)}\over2 r_+ \omega\,(m/m_*-1) } + J_m^\mathrm{low}\right\}.
\end{equation}
 \begin{equation}
 \Gamma_{\ell,m\geq 2m_*} \leq \omega\; \sinh^2\left\{  {1\over2} \ln(m/m_*-1) +  {I_{\ell, m\geq2m_*}\over2 r_+ \omega } + J_m^\mathrm{high}\right\}.
\end{equation}

\section{Discussion}\label{S:Discussion}

The net result of this article is to establish certain rigorous bounds on the greybody factors (mode dependent transmission probabilities) for Kerr--Newman black holes.  As a side effect, we have also obtained certain rigorous bounds on the emission rates for the super-radiant modes.  An interesting feature of these bounds is the ubiquity of the basic quantity $I_{\ell,m}$ which itself is simply linear in the spheroidal harmonic eigenvalue $\lambda_{\ell m}(a\omega)$. (Recall that $\lambda_{\ell m}(a\omega)\to \ell(\ell+1)$ as rotation is switched off, $a\to 0$). This seems to indicate that it is the use of separable spheroidal coordinates that is in many ways more crucial than the specific form of the metric components. 

We do not claim that these bounds are in any sense optimal. (Except, perhaps, in the restricted sense that these seem to be the easiest bounds to establish). It is quite possible that making different choices at various stages of the analysis could lead to tighter bounds, but there are no really obvious routes to guaranteeing tighter bounds. 
Possible routes to explore might include the ``Case II'' bounds of reference~\cite{bounds}, the Miller--Good version of the bounds presented in reference~\cite{miller-good}, or the general considerations of~\cite{bogoliubov, analytic, shabat-zakharov}. In a rather different direction, since transmission probabilities are intimately related to quasi-normal modes, it may prove useful to adapt the formalism and techniques of~\cite{Boonserm:2010, Boonserm:2010b, Boonserm:2011, Boonserm:2013}.

More prosaically,  there would be in principle no obstruction to adding mass and charge to the scalar field, (see for instance the Regge--Wheeler analysis in reference~\cite{Kokkotas:2010}), but the results are likely to be algebraically messy. Other possibilities to explore might include the behaviour of spin-1/2,   spin-1, and spin-2 fields, or the consideration of other interesting spacetime geometries.

\section*{Acknowledgments}

PB was supported by a grant for the professional development of new academic staff from the Ratchadapisek Somphot Fund at Chulalongkorn University, by the Thailand Toray Science Foundation (TTSF), by the Thailand Research Fund (TRF), the Office of the Higher Education Commission (OHEC), Chulalongkorn University (MRG5680171), and by the Research Strategic plan program (A1B1), Faculty of Science, Chulalongkorn University. 
\\
TN was supported by a scholarship from the Development and Promotion of Science and Technology talent project (DPST). 
\\
MV was supported by the Marsden Fund, and by a James Cook fellowship, both administered by the Royal Society of New Zealand.



\end{document}